# A Spintronic Battery with Reversible Modulation of Spin Polarization through Li Charge/Discharge: A First Principles Computational Modelling Case Study for an Antiperovskite System

*Sk Mujaffer Hossain[a,c,d], Vinila Bedekar[a], Priyanka Yadav[b], Ram Janay Chudhary\*[b], Satishchandra Ogale\*[a,c]*

[a]  Department of Physics and Centre for Energy Science, Indian Institute of Science Education and Research, Pune 411008, India

[b]  UGC-DAE Consortium for Scientific Research, Indore Centre, University Campus, Khandwa Road, Indore 452001, India

[c]  Research Institute of Sustainable Energy, TCG-CREST, Salt Lake, Kolkata 700091, India

[d]  Department of Chemistry, Indian Institute of Science Education and Research, Pune 411008, India

Email: satishogale@gmail.com,  ram@csr.ernet.in

**Abstract:**

A key notion defining the progress of the emergent fields of modern electronics, renewable energy, and smart systems is "charge storage" which is primarily embodied in various battery chemistries and systems. In addition to the "charge" property, the electron also has the "spin" property which is exploited in the field of "spintronics" to access novel magnetically controlled device actions that are not accessible to conventional electronics. An interesting question is whether the two can be fruitfully integrated into a single device concept to expand the horizon of device design and applications. Herein, we present a combined experimental and theoretical study of virgin and lithiated conducting intermetallic anti-perovskite with nominal stoichiometry represented as $Li_xFe_3SnC$ (x = 1, 2, 3, 4) to establish the principle of reversible and concurrent charge and spin polarization storage that can be aptly christened as Iono-Spintronics representing a notion of Spintronic Battery. The experimental results however showed that lithiation turns the system into a biphasic state comprised of tin-lithium alloy (due to high affinity of Sn for Li) along with lithiated $Fe_3C$; the process exhibits multiple cyclability (rechargeability).

## 1. Introduction

An electron is endowed with two important properties namely charge and spin. Manipulating the flow of electronic charge in various ways in materials media is the mandate of modern electronics and the world has witnessed its tremendous impact on our lives over the past several decades. During the past few decades, there has also been significant progress in manipulating the spin property of electrons to generate novel magnetically manipulated device architectures and responses that are conceptually entirely different from those of conventional electronics. This branch of physics, termed Spintronics, also promises interesting new solutions for our futuristic technological needs in fields such as IOT.

One of the key device constructs of modern electronics that has fueled very rapid progress of the emergent fields of renewable energy and smart mobile device systems is the concept of "charge storage" embodied in various battery systems that have been developed over many decades. During the past two decades, a question that has surfaced and been discussed in the literature in a limited way is whether a "spin battery" is feasible along the same lines as a "charge battery". A "spin battery"[1] envisages device architectures that could provide pure spin currents into the external circuits without associated charge current. These studies have involved some very creative suggestions involving electric field controlled selective spin injection across a semiconductor interface[2–4], or use of topological insulators, etc[5]. A few other studies, which have no direct bearing on the concept of "spin battery" as stated above, have addressed a separate question of manipulating the magnetic moment of some magnetic materials by electrochemical lithiation and delithiation as done in a rechargeable battery system for "charge" storage. To put our work into perspective, we briefly discuss below some of these past research works.

In an early study, Sivakumar et al. (MIT group) examined the changes in the magnetic and structural properties of magnetite ($Fe_3O_4$) as a function of lithiation. An almost 75% drop of saturation moment was observed upon 2 moles of Lithium insertion per formula unit.[6] Again in his other paper, he showed that by electrochemical lithiation a large change in the magnetization of $CrO_2$ is possible[6]. In another work, Abdel-Ghany et al. studied the magnetic properties of $LiNi_{0.33}Mn_{0.33}Co_{0.33}O_2$ (lithiated NMC) more from the standpoint of using magnetism as a diagnostic tool for developing a fundamental understanding and control of battery phenomena[7]. However, no battery electrochemistry was used to examine reversible charge-discharge effects. In a subsequent and more recent work, Reitz et al. explored the magnetism of mesoporous Lithium ferrite films and demonstrated *in situ* tuning of their magnetic properties by reversible topotactic lithium insertion[8]. In a somewhat similar study based on nanoscale hematite battery electrode, Zhang et al. showed reversible and rapid manipulation of its magnetism by 3 orders of magnitude by lithiation/delithiation at room temperature[9]. In another purely theoretical recent study, Wang et al. considered the case of half-metallic (100% spin polarization) $TiF_3$ under lithium insertion and showed that the half metallicity is retained upon insertion[10]. They proposed that such material could potentially serve as an anode for the Li-ion spin battery.

One must differentiate between the technical interest in reversibly changing the magnetic moment and/or coercive field of a ferromagnet (or a ferrimagnet or canted antiferromagnet) by lithiation/de-lithiation and changing its spin polarization, because the latter is important for spintronic devices wherein spin itinerancy is the key, rather than static information storage. Indeed, we believe that there is significant scope for doing new and novel device science if we do not restrict to the concepts of separate charge and spin storage. Indeed, one could envision a new domain of "iono-spintronics" wherein ionic movements and separations such as in Li-ion batteries could also be used to concurrently control the spin polarization of the material. The origin of the present work lies in our recent research on a 3D antiperovskite intermetallic system $Fe_3SnC$[11], wherein we found that this conducting system (itinerancy required for spintronics) is a robust high capacity anode material that can reversibly cycle four lithium ions. While exploring the consequences of lithiation and delithiation in this system by first-principles DFT calculations we discovered an interesting fact that the "spin polarization (the net difference between spin up and spin down electrons at Fermi energy)" in this "metallic" system (finite density of states at the Fermi energy $E_F$)" undergo significant changes as a function of the degree of lithiation (from 1 to 4 Li-ions). This suggested that it is possible to ionically and reversibly control the spin polarization of the anode concurrently with the charge population in an electrochemical battery configuration, a perfect case for the suggested concept of "Iono-Spintronics". Following the theoretical inputs, we experimentally examined the magnetism (along with the other properties) in the solid-state sintered extra Li-incorporated bulk system $Li_xFe_3SnC$ (x = 1, 2, 3, 4) as well as the reversible magnetism changes occurring in the same material as a function of lithiation/delithiation in a Li-ion battery half-cell. We found remarkable similarities between the theoretical and experimental results, suggesting the possible realization of a real iono-spintronic charge-spin battery effect.

3D antiperovskite is rapidly emerging as a materials system of significant interest in various fields of application; with nitride and carbide-based antiperovskite attracting the most recent attention [12,13]. The general formula for antiperovskite materials is $M_3AX$ where M is a 3D transition metal (e.g., Fe, Ni, Co, Mn, Cu, etc.), A is a post-transition metal (e.g., Sn, Ga, Ge, etc.), and (X= N, C, O). In an $M_3AX$ cubic antiperovskite material, the M atom is situated at the face center, the A atom is at the corner, and the C atom is at the body center with $M_6C$ octahedral coordination[14]. Antiperovskite material has many applications in the different fields of science and technology such as advanced battery materials[15–21], magnetism and magnetoresistance[22–27], the emergence of superconductivity[28–34], electrochemical energy conversion and storage[35–37], etc.

## 2. Computational Details

We performed spin-polarized ab initio density functional theory (DFT) based calculations using the Quantum Espresso (QE) computer package[52]. Ultrasoft pseudopotentials were used to describe the electron-ion interactions[53]. The valence electronic configurations, considered in this work, for each element are: Sn [$4d^{10}$ $5s^2$ $5p^2$], C [$2s^2$ $2p^2$], Fe [$3s^2$ $3p^6$ $3d^6$ $4s^2$], and Li [$1s^2$ $2s^1$]. A plane-wave basis set determined by kinetic energy cut-offs of 60 Ry (for wavefunction) and 520 Ry (for charge density) was used to expand the wavefunction and charge density. The electron-electron exchange-correlation was described by the Perdew-Bueke-Ernzerhof (PBE) parametrization of the generalized gradient approximation[54]. Brillouin zone integrations were performed with a shifted (12×12×12) Monkhorst-Pack[55] k-point mesh. Marzari-Vanderbilt smearing of 0.008 Ry was used to speed up the convergence[56].

## 3. Results and Discussion

### 3.1. Lithiated Antiperovskite System

$Fe_3SnC$ is a cubic antiperovskite intermetallic compound. In this crystal, eight tetrahedral and three octahedral interstitial void spaces are present, hence there is sufficient space for a Lithium atom to navigate within the cubic unit cell of $Fe_3SnC$. In our earlier paper, we showed that a maximum of 4 lithium atoms can be stored in bulk $Fe_3SnC$ through electrochemical lithiation and this material behaves as a high-capacity anode with a very impressive specific capacity of 450 mAh/g.[11] However, in that study, it could not be unambiguously discerned whether the system retains its single-phase character upon multiple lithium-ion incorporations.

Taking clues from these experimental inputs and to plan the present work, we first modeled the systematic insertion of 1-4 Lithium atoms at different interstitial positions in the bulk $Fe_3SnC$ crystal and found that there are 34 different inequivalent configurations as shown in **Figure S1**. Upon optimizing all these structures (**Figure S2**), we observed that due to the insertion of the Li atom(s) in the unit cell of $Fe_3SnC$, significant strain and structural deformations occur. The theoretically obtained lowest energy lithiated crystal structures for 1-4 Li are shown in **Figure S2** and their optimized lattice parameters are given in **Table S1.** It can be seen from **Figures (Tables S1)** and **S2** that the theory suggests that after one Li incorporation, the cubic phase of $Fe_3SnC$ must get distorted and become rhombohedral (hexagonal family) if it has to retain its single-phase character.

### 3.2. Magnetic Properties unlithiated $Fe_3SnC$

In the earlier reports, there has been a controversy about the magnetic ground state of $Fe_3SnC$. Stadelmaier and Huetter suggested that this ternary carbide is ferromagnetic [44]. Ivanovski et al. reported that this is nonmagnetic[43,45], another experimental group of Wang et al.[43] reported that it could behave as spin glass material and Grandjean reported that $Fe_3SnC$ may have an FIM coupling phase below 77 K[46]. In our DFT study, we have observed that the ferrimagnetic (FIM) order is more favorable as compared to nonmagnetic (NM) and ferromagnetic (FM) orders as the NM and FM are 2.42 and 37.5 meV higher in energy than the FIM order, respectively, as shown in **Table 1.** The arrangements of electron spin for FM and FIM orders of the Fe atoms in the lattice are shown in **Figure 1 (a, b)**. Further, from the plots of the density of states (DOS) and their contributions from the different atomic orbitals of Fe, C, and Sn (projected DOS in **Figure 1d**) we observe that not only the magnetic moments are arising primarily from the exchange splitting of the Fe-d states but also these have the most dominant contributions at the Fermi energy ($E_F$). Importantly, we also observe a large difference between the contribution to the DOS at $E_F$ from the spin-up and spin-down channels. To quantify this difference we compute the spin polarization (SP) using the following formula:[48–51]

$$Spin\,Polarization(SP) = \frac{N^{\uparrow}_{E_F} - N^{\downarrow}_{E_F}}{N^{\uparrow}_{E_F} + N^{\downarrow}_{E_F}} \qquad (1)$$

$$Degree\,of\,spin\,polarization = |SP| \times 100\% \qquad (2)$$

$N^{\uparrow}_{E_F}$ and $N^{\downarrow}_{E_F}$ are the number of spins up and spin down states, respectively, at the Fermi level ($E_F$). We find that for the pristine case *SP*=30.39%.

**Table 1.** Energy difference(ΔE) between the different magnetic orders of $Fe_3SnC$ from the lowest energy FIM ordering.

| Magnetic Order | ΔE (meV) |
|---|---|
| Non-magnetic (NM) | 2.42 |
| Ferromagnetic (FM) | 37.5 |
| Ferrimagnetic (FIM) | 0.0 |

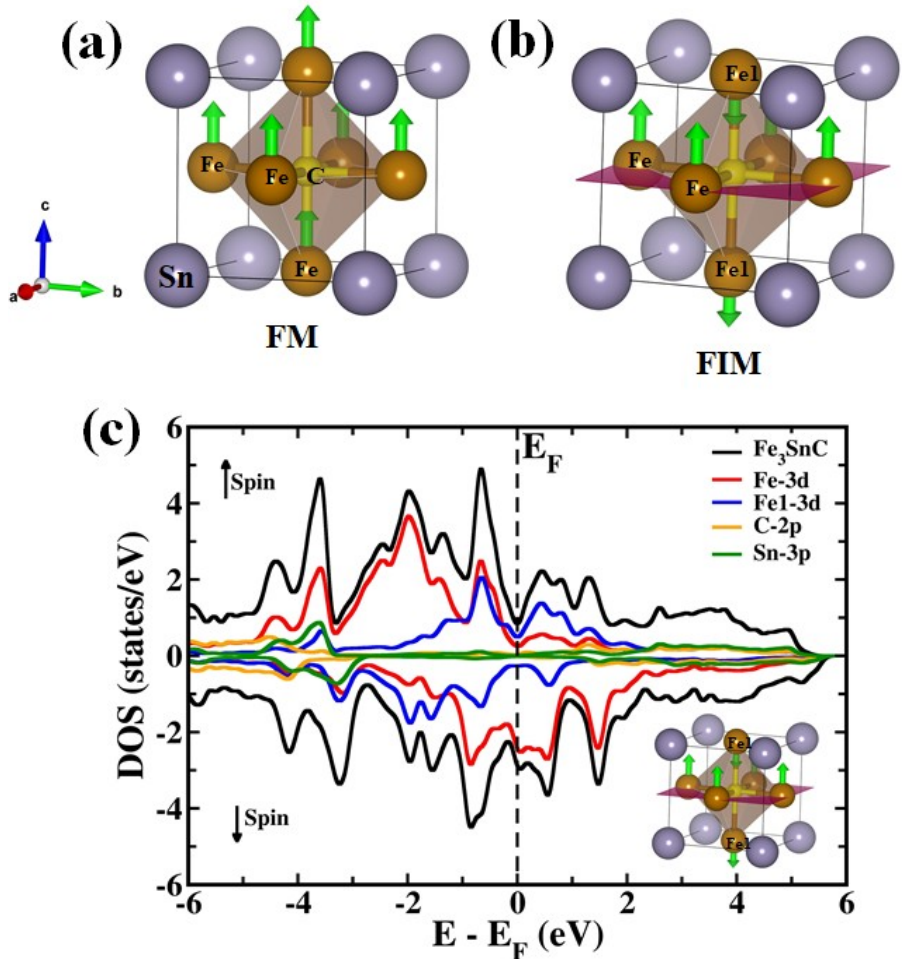

**Figure 1.** (a, b) represents the spin arrangement of the Fe atom of FM and FIM order, the green arrow represents the arrangement of the direction of the spin magnetic moment, and (c) represents the DOS of FIM of $Fe_3SnC$.

**3.2. Li induced Strain and Magnetization in $Fe_3SnC$**

In our previous theoretical study, we have already seen that the strain effects due to the lithiation indeed cause the structural changes at the time of charging and discharging of Li-ions, which are shown in the supporting information[11]. In the current study, we focus on the magnetic consequences of these effects by performing further theoretical studies.

In order to understand the effect of Li insertion on the magnetic properties and spin-polarization of $Fe_3SnC$ we have computed the same for the Li-inserted case. For these, we have considered both FM and FIM magnetic ordering **Figure 1(a, b)**. To study FIM ordering in the Li inserted systems we considered a starting configuration as shown in **Figure 1(b).** For all the cases we find that the magnetic configurations are lower in energy than the non-magnetic ones. Amongst the magnetic configurations, the energy differences between the lowest energy structures of FM and FIM order in $Li_{x=1-4}Fe_3SnC$ are given in **Table S1** of the supporting information (SI). We observe that after optimization $LiFe_3SnC$, $Li_2Fe_3SnC$, and $Li_4Fe_3SnC$ prefer FM ordering whereas $Li_3Fe_3SnC$ is stable in FIM configuration. Interestingly, we also observe that there are drastic structural changes. Not only do the crystal structure changes but also the local coordination around the Fe atoms change. For example, while for $Fe_3SnC$ the Fe atoms are in the $Sn_4C_2$ octahedron (**Figure 2a**), upon insertion of Li (till 3 Li insertion) the Fe atoms are in the $Sn_2Li_2C_2$ octahedron. For 2 and 3 Li atom insertions, in addition to the $Sn_2Li_2C_2$ octahedron, we also observe the $Sn_4CLi$ octahedron containing the Fe atoms (**Figure 2 b, c**). Both these octahedra are highly distorted. Moreover, in $Li_3Fe_3SnC$ we observe layering; three distinct layers can be observed. The spin-down-Fe and Sn form a layer that is followed by spin-up-Fe and C. This is followed by a highly buckled layer of Li atoms (**Figure 3**). The insertion of 4 Li atoms completely destroys these octahedra. We also note that for 1, 2, and 3 Li insertion, the Fe-Fe distances are reduced (between 2.43 – 2.64 Å) compared to that of $Fe_3SnC$ (2.74 Å).

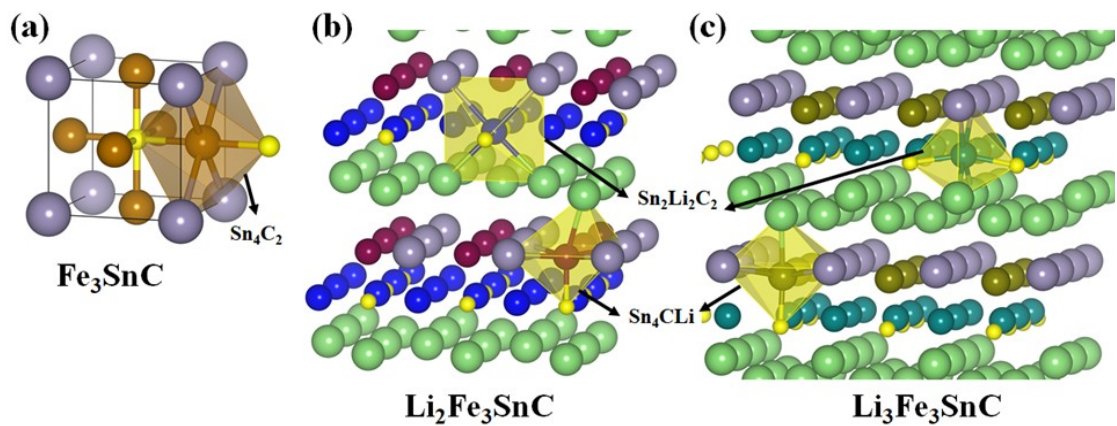

**Figure 2**. (a) $Sn_4C_2$ octahedra in $Fe_3SnC$ (b, c) $Sn_2Li_2C_2$ and $Sn_4CLi$ octahedra in $Li_2Fe_3SnC$ and $Li_3Fe_3SnC$ respectively.

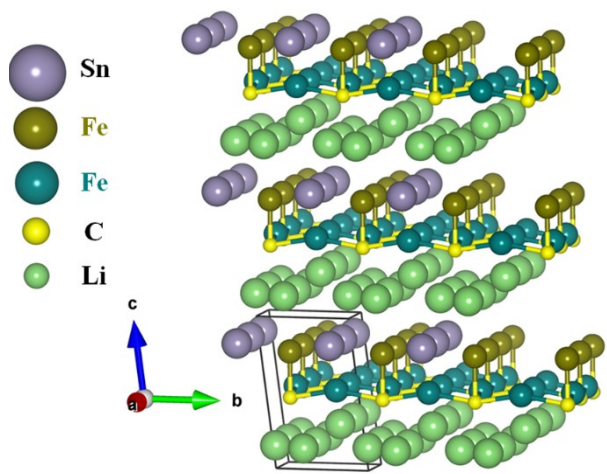

**Figure 3**. The repeating unit of $Li_3Fe_3SnC$ where **Fe** represents the down-spin and **Fe** represents the up-spin.

### 2.3. Li-induced Spin Polarization in $Fe_3SnC$

The above-mentioned structural distortions (**Figures 2 and 3**) induce significant changes in hybridization between the atomic orbitals of Fe, C, Sn, and Li, which in turn results in changes in the DOS and magnetic properties of lithiated $Fe_3SnC$. A comparison of the total DOS of the lithiated cases with pristine $Fe_3SnC$ is shown in **Figure 4.** Analogous to the pristine compound, all the lithiated cases exhibit metallic behavior. Further, in all these cases, it is the Fe-d states that have dominant contributions to the DOS at Fermi energy. More interestingly, the spin polarization at the Fermi level is significantly modulated. *SP* as a function of the number of Li atoms is shown in **Figure 9** (wine curve). We observe that except for $Li_3Fe_3SnC$ (insertion of three Li-ions), where the system is in an FIM configuration, lithium insertion significantly enhances the spin polarization at $E_F$ as shown in **Table 2**. The largest enhancement (**more than twice**) is observed for $LiFe_3SnC$.

**Table 2.** Contribution of up and down spin of electrons at the Fermi level into the magnitude of spin polarization.

| Systems | $N_\uparrow(E_F)$ | $N_\downarrow(E_F)$ | SP=$\|\frac{N^\uparrow_{E_F} - N^\downarrow_{E_F}}{N^\uparrow_{E_F} + N^\downarrow_{E_F}}\|$ x100% |
|---|---|---|---|
| **$Li_0Fe_3SnC$** | 1.28 | 2.36 | 30.39 |
| **$LiFe_3SnC$** | 0.65 | 3.87 | 71.23 |
| **$Li_2Fe_3SnC$** | 1.31 | 3.02 | 39.5 |
| **$Li_3Fe_3SnC$** | 2.60 | 2.04 | 12.09 |
| **$Li_4Fe_3SnC$** | 2.01 | 4.47 | 38.00 |

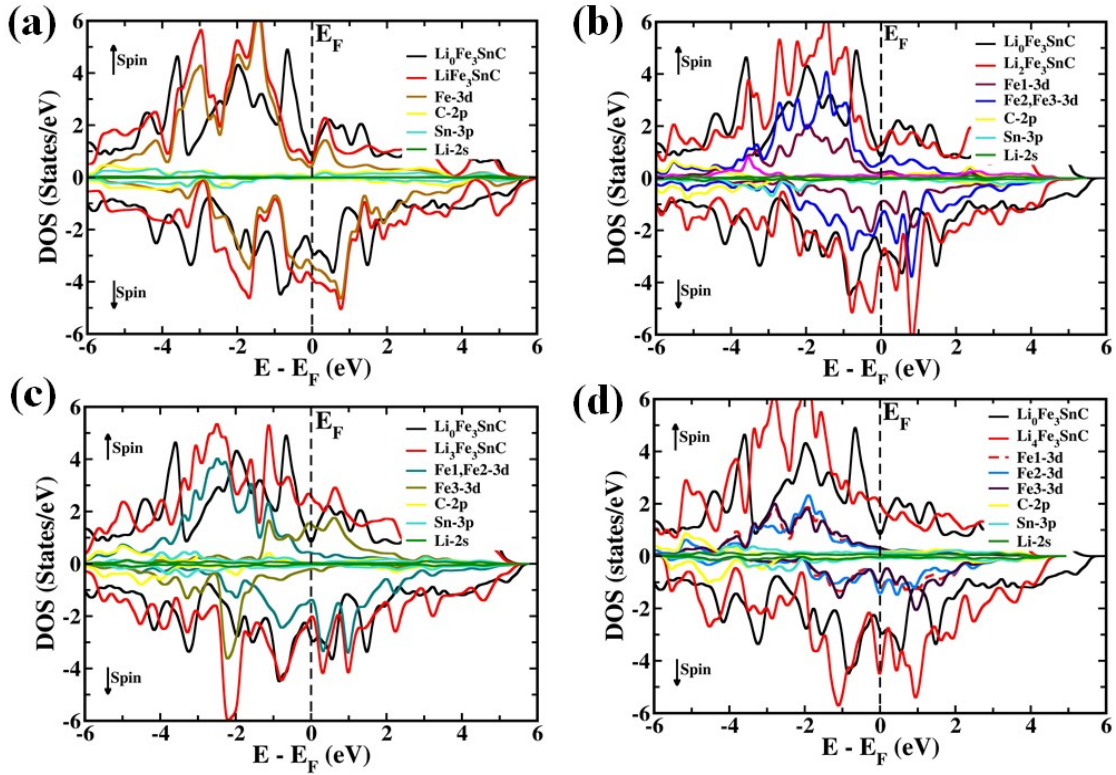

**Figure 4.** Represents the total density of states (DOS) of pure $Fe_3SnC$ (black curve) and projected DOS of lithiated system. DOS of (a) $Fe_3SnC$ (FIM), $LiFe_3SnC$ (FM) (b) $Li_2Fe_3SnC$ (FM) (c) $Li_3Fe_3SnC$ (FIM) (d) $Li_4Fe_3SnC$ (FM).

## 2.4. Formation of LiSn and $Fe_3C$ phases after first lithiation in $Fe_3SnC$

Our DFT calculations confirmed that pure $Fe_3SnC$ indeed changes its crystal system from cubic to rhombohedral after the incorporation of a single lithium-ion and form LiSn and $Fe_3C$ two phases as shown in **Figure 5c.** This happens purely due to the Coulomb repulsion between the Li ions and the atom inside $Fe_3SnC$ which induces enormous strain in the $Fe_3SnC$ lattice. The strain is released to a large extent by the lattice undergoing a structural transition from the cubic to the rhombohedral phase. As a consequence, when a Li-ion is inserted into the tetrahedral position in $Fe_3SnC$ **(Figure 5b)**, the tetrahedron is significantly distorted; while the axis along with the Sn and the center of the basal plane formed by the three Fe atoms is significantly elongated (2.22Å in $Fe_3SnC$ vs. 4.55Å in $LiFe_3SnC$ (**Figure 6**), the triangular basal plane is compressed.

As Li has a stronger tendency to react with Sn as compared to Fe[41], the Sn-Fe bond length is about 0.03 Å shorter than the Li-Fe bond length of 2.48 Å. The compression of the basal plane of the SnFe$_3$ tetrahedron is evident from the significant reduction of the Fe-Fe bond length (2.75 Å in Fe$_3$SnC vs. 2.43Å in LiFe$_3$SnC). Moreover, we also observe slight contraction of the Fe-C bond lengths (**Figure 5c**).

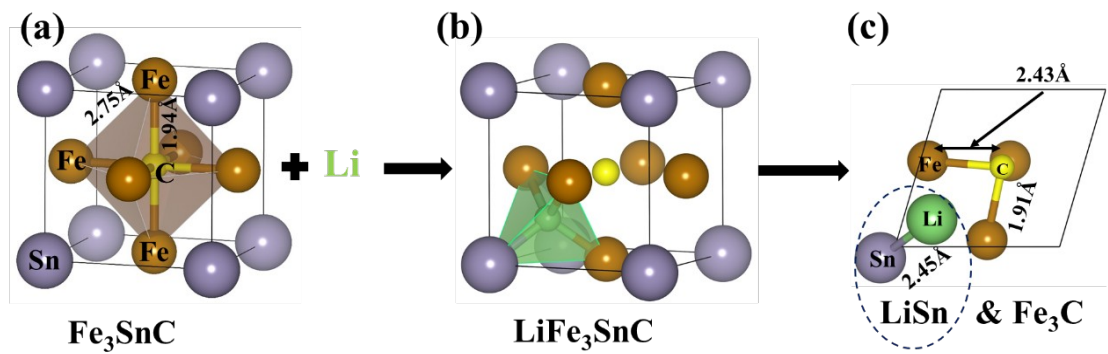

**Figure 5.** (a) Pure Fe$_3$SnC crystal system (b) Li at a tetrahedral position inside Fe$_3$SnC (c) Optimize Fe$_3$SnC with a single lithium (n=1) and formation of LiSn and Fe$_3$C phases.

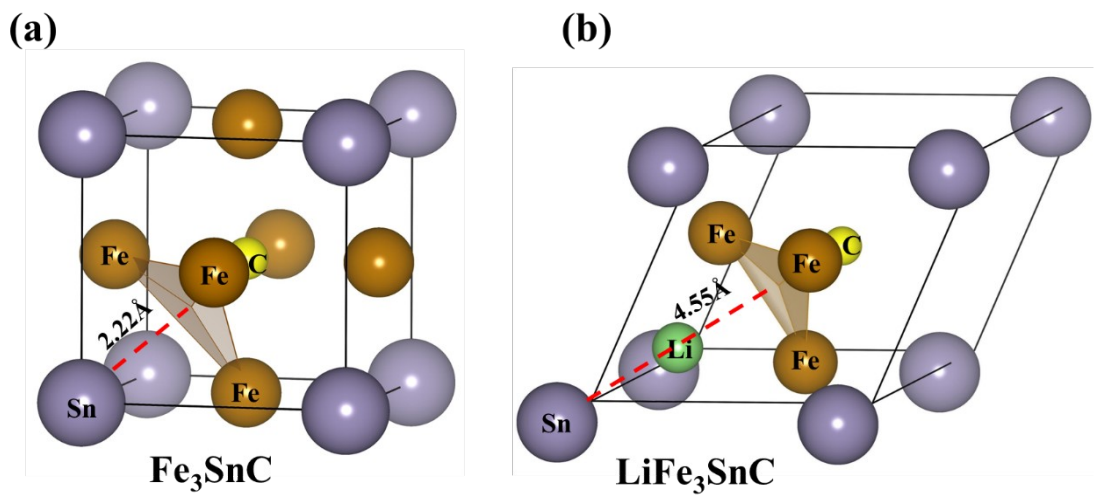

**Figure 6**. (a, b) Distance between the Sn and the center of the basal plane formed by the three Fe atoms for $Fe_3SnC$ and $LiFe_3SnC$ respectively.

## 2.5. Li-induced Spin Polarization in $Fe_3C$

Since, as we found that lithiation evolves the sample into a biphasic constitution involving a tin-lithium alloy (Sn getting extracted from the $Fe_3SnC$ phase to form the alloy) and a lithiated $Fe_3C$ phase component; we examined theoretically (DFT) whether the magnetism and the changes therein emanate from the latter phase due to lithiation-induced strain and related changes. The cubic (Pm-3m) phase of $Fe_3C$ has been taken from the material project database to perform the calculation similar to the $Fe_3SnC$ shown in **Figure S3** and the corresponding crystal lattice parameters have been presented in **Table S2.**

With lithiation, the formation of structural distortion strain/stress alters the local coordination around the Fe atom and this leads to the changes in the magnetic moments per Fe atom (**Figure. 7(a-d)**). The above-mentioned structural distortions induce significant changes in hybridization between the atomic orbitals of Fe, C, and Li, which in turn result in changes in the DOS and magnetic properties of lithiated $Fe_3C$ shown in **Figure 8**.

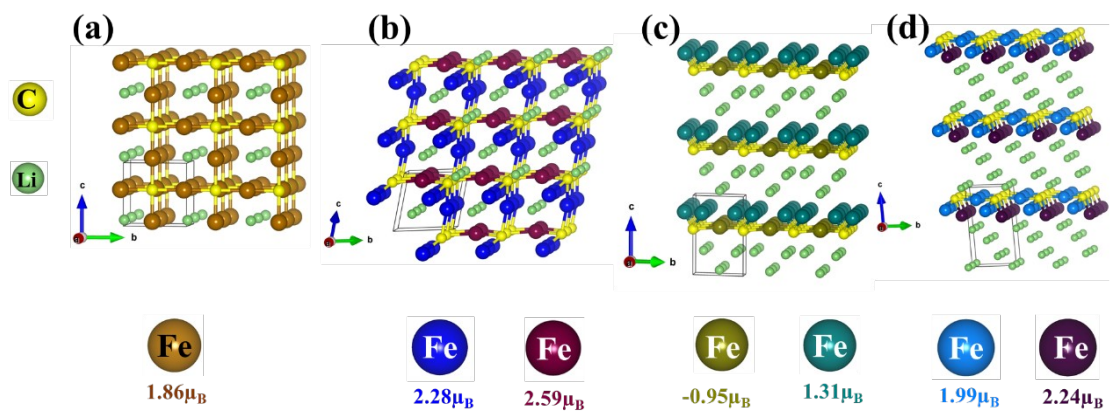

**Figure 7.** The repeating unit of lithiated phases of Fe$_3$C: (a) LiFe$_3$C (b) Li$_2$Fe$_3$C, (c) Li$_3$Fe$_3$C, and (d) Li$_4$Fe$_3$C. The different colors of the Fe atom correspond to the different magnetic moments

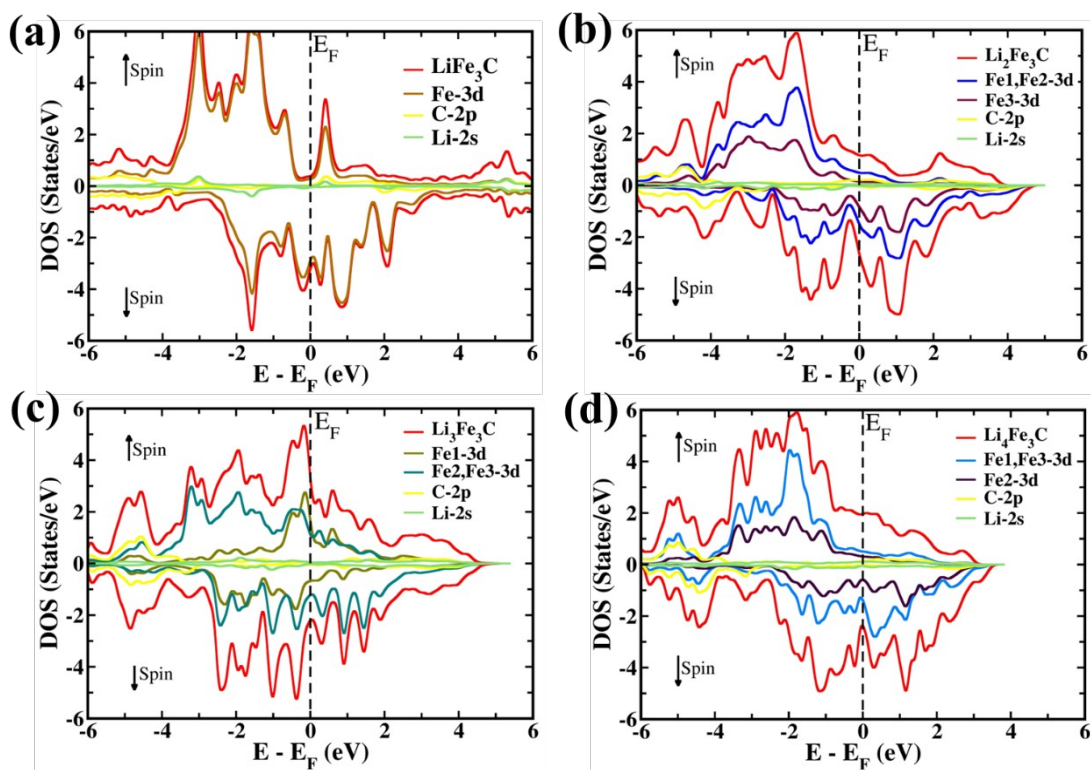

**Figure 8.** Represents the total density of states (DOS) of pure Fe$_3$C (black curve) and projected DOS of lithiated system, (a) Fe$_3$C (FM), LiFe$_3$C (FM) (b) Li$_2$Fe$_3$C (FM) (c) Li$_3$Fe$_3$C (FIM) (d) Li$_4$Fe$_3$C (FM).

As seen from the data shown in **Fig. 8**, all the lithiated cases of Fe$_3$C exhibit metallic behavior similar to the unlithiated Fe$_3$SnC and Fe$_3$C. Further, in all these cases, it is the Fe-d states that have dominant contributions to the DOS at the Fermi energy. More interestingly, the spin polarization at the Fermi level is significantly modulated. *SP* as a function of the number of Li atoms is shown in **Fig. 9** (red curve). We observe that except for Li$_3$Fe$_3$C (insertion of three Li-ions), where the system is in an FIM configuration (**Fig. 6c**), lithium insertion significantly enhances the spin polarization at E$_F$ as shown in **Tables 2** and **3**. Noting that for the pristine cases of Fe$_3$SnC and Fe$_3$C the *SP* values are obtained as 30.39% and 13.64%, respectively.

**Table 3.** Contribution of up and down spin of electrons at the Fermi level into the magnitude of spin polarization. (Li$_x$Fe$_3$C)

| Systems | $N_\uparrow(E_F)$ | $N_\downarrow(E_F)$ | SP=$\|\frac{N^\uparrow_{E_F}-N^\downarrow_{E_F}}{N^\uparrow_{E_F}+N^\downarrow_{E_F}}\|$ x100% |
|---|---|---|---|
| **Li$_0$Fe$_3$C** | 2.24 | 1.71 | 13.64 |
| **LiFe$_3$C** | 0.39 | 3.21 | 78.53 |
| **Li$_2$Fe$_3$C** | 1.16 | 2.64 | 39.01 |
| **Li$_3$Fe$_3$C** | 2.39 | 1.96 | 10.0 |
| **Li$_4$Fe$_3$C** | 2.21 | 3.34 | 20.00 |

We surprisingly observed that the spin polarization almost follows a similar trend as in the case of Fe$_3$SnC shown in **Figure 9**.

To corroborate our theoretical observations we performed some experimental analysis after synthesizing lithiated $Fe_3SnC$ (discussed in SI) and we observed that the magnetization or magnetic moment of Fe atom of all the lithiated $Fe_3SnC$ follows a similar trend of spin polarization for both the lithiated $Fe_3C$ and $Fe_3SnC$ system as shown in **Figure 9**.

Moreover, in **Figure 9**, we have also plotted the average magnetic moment per Fe atom as a function of the number of Li atoms, both computed and experimentally measured (see SI), along with the spin polarization. We note that the computed and experimentally measured values of the magnetic moments are in excellent agreement. Further, the trends in the spin polarization as a function of the number of Li atoms follow the trends in the magnetic moments on the Fe atoms.

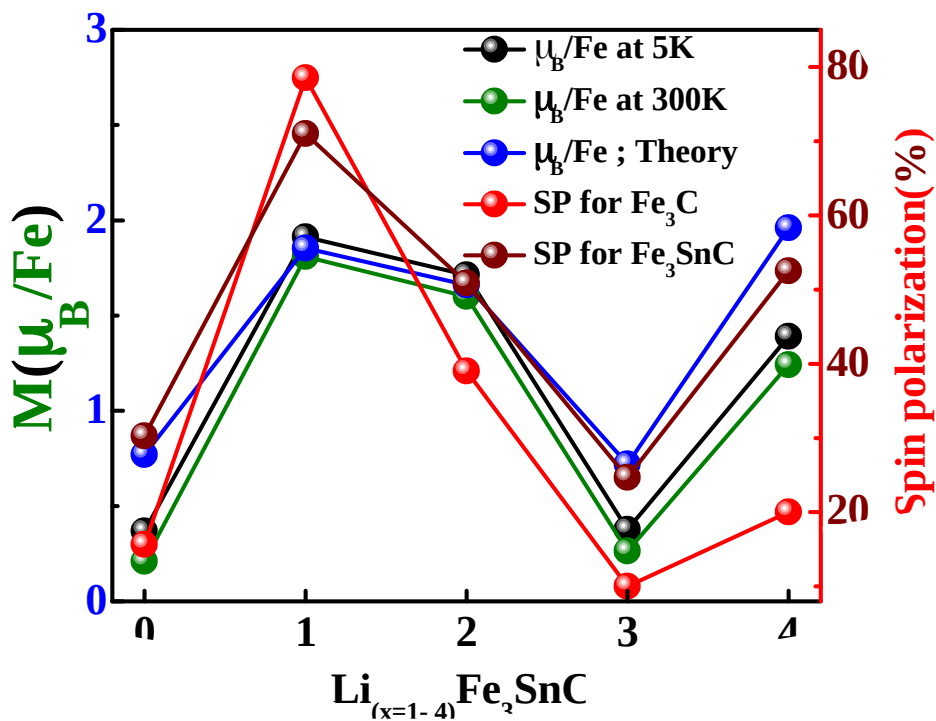

**Figure 9.** The Black and green curves represent the experimental magnetization per Fe atom at 5K and 300K, and the blue and cyan curve represents theoretically obtained magnetic moment per Fe atom for $Fe_3SnC$ and $Fe_3C$, the red and wine color curve represents the magnitude of the degree of spin polarization for both the system.

## 3. Conclusions

In this combined theoretical and experimental study, we introduce the concept of "iono-spintronics" or Spintronic Battery wherein the accumulation and de-accumulation of Li-ion in the antiperovskite $Fe_3SnC$ anode of a Li-ion half-cell concurrently and reversibly stores charge and spin; the latter mimicked by the experimental observation of non-monotonic changes in magnetization as a result of degree of lithiation, nominally depicted as ($Li_{x=1-4}Fe_3SnC$). It is noted experimentally that lithiation leads to a biphasic system in this particular case because of the affinity of Sn to form an alloy with Li. Thus with progressive lithiation changes occur in the stoichiometries and properties of the two phases, namely the Sn-Li alloy phase and lithiated or partially lithiated $Fe_3C$ phase. The computer (DFT) calculations have been performed accordingly to account for these changes. Since the system has been seen to perform as a robust anode material under multiple cycling, it seems that the non-magnetic tin-lithium alloy serves as a metallic glue that holds the anode against pulverization due to the stress effects of lithiation-delithiation.

Given the structural and strain-related changes in magnetization upon lithiation and its degree, and their connection to spin polarization outlined here, we believe that this study would have broader and more interesting implications for a variety of structurally interesting materials systems, not limited only to the anti-perovskite intermetallic case discussed herein, wherein the structure-magnetism-spin polarization connection is acute. Also, the biphasic character encountered here may not occur in other systems wherein selective preferred interaction, such as that of Sn in the present case, is not a possibility. With the rapid evolution of the fields such as the internet of things (IoT) a new breed of low-power consuming spintronics devices with novel functionalities is desired. Towards this end, effects such as concurrent storage of spins mediated by traditional battery architecture described here could add an interesting dimension.

**Supporting Information**

Supporting Information is available from the Wiley Online Library or from the author.

**Acknowledgments**

S.M.H. would like to thank IISER Pune for Ph.D. fellowships, and all the authors will like to thank DST CERI, DST Nanomission (Thematic unit program), and the Indo-UK SUNRISE program for funding support. We acknowledge the National Supercomputing Mission (NSM) for providing computing resources of 'PARAM Brahma' at IISER Pune, which is implemented by C-DAC and supported by the Ministry of Electronics and Information Technology (MeitY) and Department of Science and Technology (DST), Government of India.

**Conflict of interest**

The authors declare no conflict of interest.